\documentstyle[preprint,aps]{revtex}
\tightenlines
\begin{document}
\title{Identities involving elementary symmetric functions} 
\author{S. Chaturvedi\thanks{e-mail:scsp@uohyd.ernet.in} \\
School of Physics,\\ University of Hyderabad,\\ Hyderabad 500 046 India\\
V Gupta \thanks{:virendra@kin.cieamer.conacyt.mx}\\ 
Departamento de F\'isica Aplicado
\\Centro de 
Investigaci\'on y de Estudios Avanzados del IPN,\\ 
Unidad M\'erida, A. P. 73, Cordemex 97310,
 M\'erida, Yucatan, Mexico} 
\maketitle
\begin{abstract}
A systematic procedure for generating certain identities involving elementary 
symmetric functions is proposed. These identities, as particular cases, 
lead to new identities for binomial and q-binomial coefficients. 
\end{abstract}
\newpage
Ever since the advent of Calogero-Sutherland models$^{1-4}$ there 
has been a considerable interest in finding 
homogeneous symmetric polynomials $P_k (x) ~;~ x\equiv (x_1, x_2, \cdots,x_N )$ 
of degree $k$ which satisfy the generalized Lapace's equation  
\begin{equation}
\left[\sum_{i=1}^{N}   \frac{{\partial}^2}{\partial x_{i}^{2}} 
+\frac{2}{\alpha}
\sum_{i< j} \frac{1}{(x_i-x_j)}\left( \frac{\partial}{\partial x_i}
-\frac{\partial}{\partial x_j}\right) \right]P_k (x) = 0~~~.
\label{gle}
\end{equation}
Since one is seeking solutions to (\ref{gle}) which are symmetric functions 
of $(x_1, x_2,\cdots, x_N)$ it appears natural to change variables from 
$(x_1, x_2,\cdots)$ to a set of variables which are symmetric functions 
of $(x_1, x_2,\cdots)$ and rewrite the 
generalized Laplace's equation in terms of these variables. Two 
sets of such variables that have been considered in the literature$^{5,6}$
repectively  are 
\begin{itemize}
\item power sums: 
\begin{equation}
p_r(x) = \sum_{i} x_{i}^{r}~~;~ r=1,\cdots,N \,\,\,\,.
\end{equation}

\item elementary symmetric functions:
\begin{equation}
e_r(x) = \sum_{i_1<i_2\cdots\ <i_r} x_{i_1}x_{i_2} \cdots x_{i_r}~;~
i_1, \cdots, i_r = 1, \cdots, N~;~ r=1, \cdots, N.
\end{equation}
\end{itemize}     
(Here, for symmetric functions, we follow the nomenclature and notation of 
ref 7) Explicit expressions for the generalized Laplace's equation in terms 
of these variables may be found in refs 5 and 6 respectively. The next step 
consists in finding polynomial solutions of the equation thus obtained.  
( It may be noted here that a more efficient way of constructing the 
symmetric polynomial solutions of (\ref{gle}) based on expanding $P_k (x)$ 
in terms of Jack polynomials$^{8}$ may be found in ref 9.) 

In changing variables from $(x_1, \cdots, x_N)$ to $(e_1(x), \cdots, e_N (x))$ 
in the generalized Laplace's equation, in the intermediate stages, 
one needs to express the symmetric function 
\begin{equation}
\sum_{i} e_{p-1}^{(i)} (x) e_{q-1}^{(i)} (x)
\label{sym}
\end{equation} 
in terms of the $e_{r}(x)$. Here $e_{p}^{(i)} (x) $ denotes the $p^{th}$ 
elementary symmetric function formed from $(x_1, \cdots, x_N)$ omitting 
$x_i $. The purpose of this letter is to provide a derivation of the 
expression of the symmetric function in (\ref{sym}) in terms of the 
elementary symmetric functions in the full set of variables 
$(x_1, \cdots, x_N )$. The procedure adopted for deriving this result 
permits easy extension to symmetric functions like 
\begin{equation}
\sum_{i=1}^{N} e_{p-1}^{(i)} (x) e_{q-1}^{(i)} (x)e_{r-1}^{(i)} (x)
\end{equation} 
and so on. Further, on setting $x_1 =\cdots=x_N =1$ in these relations, 
( $x_1 =1, x_2 = q,\cdots, x_N =q^{N-1}$) one is led to a series of  
interesting nonlinear identities for binomial (q-binomial) 
coefficients. 

To obtain the desired results, it proves convenient to work with 
the generating function for the elemantary symmetric functions 
\begin{eqnarray}
E(x,t) &=& \sum_{r=0}^{\infty} t^r e_{r} (x)\\
&= &\prod_{i=1}^{N} (1+x_i t)
\end{eqnarray}
with $e_0 (x)$ taken to be equal to $1$. From the definition of $E(x,t)$, it 
follows that 
\begin{equation}
\sum_{i=1}^{N}\frac{\partial}{\partial x_{i}}\log E(x,t)=t\sum_{i=1}^{N}
\frac{1}{(1+x_i t)}
\end{equation}
\begin{equation}
t\frac{\partial}{\partial t}\log E(x,t)=\sum_{i=1}^{N}
\frac{x_i t}{(1+x_i t)} = N - \sum_{i=1}^{N} \frac{1}{(1+x_i t)}
\end{equation}
From these relations it follows that
\begin{equation}
\sum_{i=1}^{N}\frac{\partial}{\partial x_{i}}\log E(x,t) = Nt -  
t^2\frac{\partial}{\partial t}\log E(x,t)
\end{equation}
and hence
\begin{equation}
\sum_{i=1}^{N}\frac{\partial}{\partial x_{i}} E(x,t) = Nt E(x,t) -  
t^2\frac{\partial}{\partial t} E(x,t)
\end{equation}
On substituting for $E(x,t)$ from (6) and equating like powers of $t$ on 
both sides on obtains
\begin{equation}
\sum_{i=1}^{N} e_{p-1}^{(i)} (x) = (N-p+1) e_{p-1} (x)
\end{equation}
which on setting $x_1 =\cdots=x_N =1$ and using 
\begin{equation} 
e_{r}(1,1,\cdots,1) =\left( \begin{array}{c} N \\ r\end{array} \right)~~;~~
e_{r}^{(i)}(1,1,\cdots,1) = \left( \begin{array}{c} N-1 \\ r\end{array} \right)
\end{equation}
yields  
\begin{equation}
N \left( \begin{array}{c} N-1 \\ p-1\end{array} \right)
= (N-p+1)\left( \begin{array}{c} N \\ p-1\end{array} \right)
\end{equation}

Further, setting $x_1 =1, x_2 = q,\cdots, x_N =q^{N-1}$ and using 
\begin{equation}
e_{p}(1, q, \cdots, q^{N-1})= q^{p(p-1)/2}
\left[ \begin{array}{c} N \\ p\end{array} \right]
\end{equation}
and 
\begin{equation}
e_{p-1}^{(i)}(1, q, \cdots, q^{N-1})
= q^{(p-1)(p-2)/2}\sum_{u=0}^{p-1}
q^{u(u-(p-i-1)}
\left[ \begin{array}{c} N-i \\ u\end{array} \right]
\left[ \begin{array}{c} i-1 \\ p-1-u\end{array} \right]
\end{equation}
we obtain
\begin{equation}
\sum_{i=1}^{N}\sum_{u=0}^{p-1}
q^{u(u-(p-i-1))}
\left[ \begin{array}{c} N-i \\ u\end{array} \right]
\left[ \begin{array}{c} i-1 \\ p-1-u\end{array} \right]
= (N-p+1)\left[ \begin{array}{c} N \\ p-1\end{array} \right]
\end{equation}
Here 
\begin{equation}
\left[ \begin{array}{c} N \\ p\end{array} \right]
= \frac{[N]!}{[N-p]![p]!}~~;~~[N]! \equiv [N][N-1]\cdots[1]~~;~~
[N]\equiv\frac{(1-q^N)}{(1-q)}
\end{equation}
denotes the q-binomial coefficient$^{10}$. 
Omitting the points in the double summation on the lhs of (17) where the 
summand vanishes identically and changing $p-1$ to $p$, we can rewrite 
(17) as 
\begin{equation}
\sum_{i=p+1}^{N}\sum_{u=0}^{p}
q^{u(i-p)}
\left[ \begin{array}{c} N+u-i \\ u\end{array} \right]
\left[ \begin{array}{c} i-u-1 \\ p-u\end{array} \right]
= (N-p)\left[ \begin{array}{c} N \\ p\end{array} \right]
\end{equation}
For $q=1$, (17) reduces to (14) as can easily be verified.

The same strategy as above can be adopted for deriving a host of similar 
but more 
complicated identities involving elementary symmetric functions and hence 
those involving q-binomial and binomial coefficients as  
is shown below.    
\begin{equation}
\sum_{i=1}^{N}\frac{\partial}{\partial x_{i}}\log E(x,t_1)
\frac{\partial}{\partial x_{i}}\log E(x,t_2)=
t_1 t_2 \sum_{i=1}^{N}\frac{1}{(1+x_i t_1)(1+x_i t_2)}
\end{equation}
As before, we now try to express the rhs of (20) as a linear combination 
of derivatives of $\log E(x,t)$ with respect to $t$. This can be done 
using the following relation 
\begin{equation}
\frac{1}{(1+x_i t_1)(1+x_i t_2)} = 1-\frac{1}{(t_1 -t_2)}
\left[ t_{1}^2 \frac{x_i }{(1+x_i t_1)} -
t_{2}^2 \frac{x_it_2}{(1+x_i t_2)}\right]
\end{equation}
which on summing over $i$ and using (9) gives
\begin{equation}
\sum_{i=1}^{N}\frac{1}{(1+x_i t_1)(1+x_i t_2)} = 
N-\frac{1}{(t_1 -t_2)}
\left[ t_{1}^2 \frac{\partial}{\partial t_1}\log E(x,t_1)
-t_{2}^2 \frac{\partial}{\partial t_2}\log E(x,t_2)\right]
\end{equation}
Using this in the rhs of (20) one obtains
\begin{eqnarray}
\sum_{i=1}^{N}\frac{\partial}{\partial x_{i}}\log E(x,t_1)
\frac{\partial}{\partial x_{i}}\log &E&(x,t_2) =  
Nt_1 t_2 \nonumber\\ 
&-&\left(\frac{t_1 t_2}{t_1-t_2}\right) 
\left[ t_{1}^{2} \frac{\partial}{\partial t_1}\log E(x,t_1)
- t_{2}^{2} \frac{\partial}{\partial t_2} \log E(x,t_2) \right]
\end{eqnarray}  
or 
\begin{eqnarray}
\sum_{i=1}^{N}\frac{\partial}{\partial x_{i}}E(&x&,t_1)
\frac{\partial}{\partial x_{i}}E(x,t_2) =  
Nt_1 t_2 E(x,t_1) E(x,t_2) \nonumber\\ &-& 
\left(\frac{t_1 t_2}{t_1-t_2}\right) 
\left[ t_{1}^{2} \left(\frac{\partial}{\partial t_1} E(x,t_1)\right) E(x,t_2)
- t_{2}^{2} E(x,t_1)\left(\frac{\partial}{\partial t_2} E(x,t_2)\right) \right]
\end{eqnarray}  

On substituting from (6) and equating like powers of $t_1$ and $t_2$ on 
both sides one obtains 
\begin{equation}
\sum_{i=1}^{N} e_{p-1}^{(i)} (x) e_{q-1}^{(i)} (x) 
= (N-p+1)e_{p-1}(x)e_{q-1} (x)
-\sum_{r=0}^{q-2} (p+q-2-2r) e_{p+q-2-r} (x) e_{r} (x)
\end{equation}
which is the desired result valid for $p \geq q\geq 2$.

Setting $x_1, \cdots, x_N =1$ and using (13)
one obtains the following identity 
\begin{eqnarray}
N\left( \begin{array}{c} N-1 \\ p-1\end{array} \right)
\left( \begin{array}{c} N-1 \\ q-1\end{array} \right) 
&=& (N-p+1)\left( \begin{array}{c} N \\ p-1\end{array} \right)
\left( \begin{array}{c} N \\ q-1\end{array} \right)\nonumber\\
&-& \sum_{r=0}^{q-2} (p+q-2-2r)
\left( \begin{array}{c} N \\ p+q-2-r\end{array} \right)
\left( \begin{array}{c} N \\ r\end{array} \right)
\end{eqnarray}
On rearranging the terms this identity may be rewritten as follows 
\begin{eqnarray}
&&\left( \begin{array}{c} N-1 \\ p-1\end{array} \right)\left[
\left( \begin{array}{c} N \\ q-1\end{array} \right) 
-\left( \begin{array}{c} N-1 \\ q-1\end{array} \right)\right]\nonumber\\
&= & \sum_{r=0}^{q-2} 
\left( \begin{array}{c} N-1 \\ p+q-3-r\end{array} \right)
\left( \begin{array}{c} N \\ r\end{array} \right)
- \sum_{r=1}^{q-2} 
\left( \begin{array}{c} N \\ p+q-2-r\end{array} \right)
\left( \begin{array}{c} N-1 \\ r-1\end{array} \right)
\end{eqnarray}
On using the relation
\begin{equation}
\left( \begin{array}{c} N \\ q-1\end{array} \right) 
-\left( \begin{array}{c} N-1 \\ q-1\end{array} \right)
=\left( \begin{array}{c} N \\ q-2\end{array} \right) 
\end{equation}
and making the replacements $N\rightarrow N+1, p\rightarrow p+1, 
q\rightarrow q+2$, and rearranging one obtains 
\begin{eqnarray}
\left( \begin{array}{c} N \\ p-1 \end{array} \right)
\left( \begin{array}{c} N \\ q\end{array} \right)&=&
\sum_{s=0}^{q} \left[\left( \begin{array}{c} N+1 \\ p+q-s\end{array} \right)
\left( \begin{array}{c} N \\ s\end{array} \right)
- \left( \begin{array}{c} N \\ p+q-s\end{array} \right)
\left( \begin{array}{c} N+1 \\ s\end{array} \right)\right]
\end{eqnarray}
vaild for $p \geq q$.

The basic strategy for deriving higher identities should now be clear. To 
express (5) in terms of elementary symmetric functions, one needs to consider 
\begin{equation}
\sum_{i=1}^{N}\frac{\partial}{\partial x_{i}}\log E(x,t_1)
\frac{\partial}{\partial x_{i}}\log E(x,t_2)
\frac{\partial}{\partial x_{i}}\log E(x,t_3)
=t_1 t_2 t_3\sum_{i=1}^{N}\frac{1}{(1+x_i t_1)(1+x_i t_2)
(1+x_i t_3)} 
\end{equation}
The next step consists in expressing 
\begin{equation}
\frac{1}{(1+x_i t_1)(1+x_i t_2)(1+x_i t_3)}
\end{equation}
as
\begin{equation}
\frac{1}{(1+x_i t_1)(1+x_i t_2)(1+x_i t_3)} = 
1+f_1 \frac{x_i }{(1+x_i t_1)} +
f_2 \frac{x_i }{(1+x_i t_2)}+f_3 \frac{x_i }{(1+x_i t_3)}
\end{equation}
where the $f_i$'s are functions of $t_i$'s only. This can always be done. 
This relation, in turn, allows one to express 
the rhs of (30) as a linear combination of derivatives of $\log E(x,t)$ 
with respect to $t$ and hence leading to the identities of the type 
discussed above.  Note that to derive the identities for the binomial 
coefficients alone one could have set all $x_i$'s equal to $x$ from the very 
outset.  The systematic procedure outlined here leads to much more general 
results from which the binomial identities and q-binomial identities 
arise as special cases.

\end{document}